\documentclass[letter]{aa}  

\usepackage{graphicx}
\usepackage{txfonts}
\usepackage{lscape}
\usepackage{placeins}
\usepackage[colorlinks=true,citecolor=blue,linkcolor=blue,urlcolor=blue]{hyperref}
\begin{document}

\title{The importance of radiative pumping on the emission 
     of the H$_2$O submillimeter lines in galaxies}

   \author{Eduardo Gonz\'alez-Alfonso
          \inst{1}
          \and
          Jacqueline Fischer
          \inst{2}
          \and
          Javier R. Goicoechea
          \inst{3}
          \and
          Chentao Yang
          \inst{4}
          \and
          Miguel Pereira-Santaella
          \inst{5}
          \and
          Kenneth P. Stewart
          \inst{2}
          }

   \institute{Universidad de Alcal\'a, Departamento de F\'{\i}sica
     y Matem\'aticas, Campus Universitario, E-28871 Alcal\'a de Henares,
     Madrid, Spain\\
              \email{eduardo.gonzalez@uah.es}
              \and
             George Mason University, Department of Physics \& Astronomy,
             MS 3F3, 4400 University Drive, Fairfax, VA 22030, USA
              \and
              Instituto de F\'{\i}sica Fundamental (IFF), CSIC.
              Calle Serrano 121-123, 28006, Madrid, Spain
              \and
              Department of Earth and Space Sciences, Chalmers University of
              Technology, Onsala Observatory, 439 94 Onsala, Sweden
              \and
              Centro de Astrobiolog\'{\i}a (CSIC-INTA), Ctra. de Ajalvir,
              km. 4, 28850 Torrej\'on de Ardoz, Madrid, Spain
             }

   \authorrunning{Gonz\'alez-Alfonso et al.}
   \titlerunning{The importance of radiative pumping on the emission 
     of the H$_2$O submillimeter lines in galaxies}
   

 
   \abstract{H$_2$O submillimeter emission is a powerful
     diagnostic of the molecular interstellar medium in a variety of sources,
     including low- and high-mass star forming regions of the Milky
     Way, and from local to high redshift galaxies.
     However, the excitation mechanism of these lines in galaxies has been
     debated, preventing a basic consensus on the physical information that
     H$_2$O provides. 
     Both radiative pumping due to H$_2$O absorption of
     far-infrared photons emitted by dust and collisional excitation
     in dense shocked gas have been proposed to explain the H$_2$O emission.
     Here we propose two basic diagnostics to distinguish
     between the two mechanisms:
     1) in shock excited regions, the ortho-H$_2$O\,$3_{21}-2_{12}$\,75\,$\mu$m
     and the para-H$_2$O\,$2_{20}-1_{11}$\,101\,$\mu$m rotational lines are
     expected to be in emission while,
     if radiative pumping dominates, both far-infrared lines are expected
     to be in absorption; 2) based on statistical equilibrium of H$_2$O
     level populations, the radiative pumping scenario predicts that the
     apparent isotropic net rate of 
     far-infrared absorption in the $3_{21}\leftarrow2_{12}$ (75\,$\mu$m)
     and $2_{20}\leftarrow1_{11}$ (101\,$\mu$m) lines should be higher than or
     equal to the apparent isotropic net rate of submillimeter emission in the
     $3_{21}\rightarrow3_{12}$
     (1163\,GHz) and $2_{20}\rightarrow2_{11}$ (1229\,GHz) lines, respectively.
     Applying both criteria to all 16 galaxies and several galactic
     high-mass star-forming regions
     where the H$_2$O\,75\,$\mu$m and 
     submillimeter lines have been observed with
     {\it Herschel}/PACS and SPIRE, we show that in
     most (extra)galactic sources the H$_2$O
     submillimeter line excitation is dominated by far-infrared pumping,
     with collisional excitation of the
     low-excitation levels in some of them. 
     Based on this finding, we revisit the interpretation of the correlation
     between the luminosity of the H$_2$O\,988\,GHz line and the source
     luminosity in the combined galactic and extragalactic sample.
}


   \maketitle

\section{Introduction}
\label{intro}

Soon after the launch of the {\it Herschel Space Observatory} \citep{pil10},
spectroscopic observations with the SPIRE instrument \citep{gri10}
revealed strong H$_2$O submillimeter (hereafter submm) emission from the
ultraluminous infrared galaxy (ULIRG) Mrk\,231, with strengths comparable
to the those of the CO lines \citep{wer10}. The H$_2$O submm lines were
subsequently detected in a variety of local galaxies with SPIRE
\citep[see][and references therein]{yan13,lu17} and also HIFI \citep{liu17},
as well as in high-redshift ULIRGs and Hyper-LIRGs from ground based facilities
\citep[e.g.][]{omo11,omo13,wer11,yan16,yan19}. Based on the H$_2$O line fluxes
in galaxies at all redshifts, \cite{omo13} and \cite{yan13} found a strong
correlation between the H$_2$O line luminosities and the infrared luminosities
of the galaxies, $L_{\mathrm{H_2O-submm}}\propto L_{\mathrm{IR}}^{\alpha}$,
with an index $\alpha$ slightly higher than unity. The extragalactic
H$_2$O submm emission has been modeled in terms of radiative pumping
\citep{gon10,gon14,gon21,per17},
and the $L_{\mathrm{H_2O-submm}}-L_{\mathrm{IR}}$ correlations appeared to support
this view \citep{omo13,yan13,lu17}.

In parallel, several H$_2$O submm lines have been observed in a
large number of both low-mass and
high-mass star-forming regions (LMSFR and HMSFRs) in our galaxy
\citep[e.g.][and references therein]{vd21} with
the HIFI instrument \citep{gra10}. The lines were resolved with the
high spectral resolution provided by HIFI, showing broad profiles
characteristic of outflows and their associated dense shocked gas where the
H$_2$O submm emission is collisionally excited.
The H$_2$O line profiles also often showed a narrower spike at central
velocities, which contributes negligibly (often in absorption) to the
integrated emission in LMSFRs \citep[e.g.][]{kri12},
but accounts for an average of $\sim40$\%
in HMSFRs \citep{sanjose16} and is
attributed to the warm, massive envelopes around the luminosity sources.
Considering both galactic and extragalactic sources, \cite{sanjose16}
found a strong $L_{\mathrm{H_2O-submm}}-L_{\mathrm{IR}}$ correlation 
with an index $\alpha$ that was higher (and closer to unity) than when only
the galactic sources were fitted.

Using {\it Infrared Space Observatory} \citep[{\it ISO},][]{kes96}
  data (with $80''$ aperture) and radiative transfer models,
\cite{cer06} found that much of the H$_2$O
line excitation in the Orion KL outflow is driven
by the dust continuum radiation.
On the other hand, \cite{vd21} argued, based on the nearly linear
  ($\alpha=0.95\pm0.02$) correlation they found between {\it Herschel}
  H$_2$O $2_{02}-1_{11}$\,988\,GHz line luminosities of both galactic and
  extragalactic sources versus source luminosity, 
that the H$_2$O submm emission in galaxies may be a scaled-up
version of galactic sources,
where the $L_{\mathrm{H_2O-submm}}-L_{\mathrm{IR}}$ correlation is set by
  H$_2$O collisional excitation in dense shocks.

In this Letter we address this lack of consensus on the excitation
mechanism of these lines in galaxies, and thus on the physical information that
H$_2$O provides on the sources from which the H$_2$O emission lines arise. 
We propose both a qualitative and a quantitative diagnostic
to distinguish between the two mechanisms,
each involving the far-infrared (far-IR) H$_2$O $3_{21}-2_{12}$\,75\,$\mu$m
  and $2_{20}-1_{11}$\,101\,$\mu$m lines,
and apply them to all extragalactic
sources and a sample of HMSFRs where the H$_2$O\,75\,$\mu$m
and 1163\,GHz lines have been observed with {\it Herschel}/PACS
  \citep{pog10} and SPIRE, respectively.
A flat $\Lambda$CDM cosmology with
  $H_0=70$\,km\,s$^{-1}$\,Mpc$^{-1}$ and
  $\Omega_{\mathrm{M}} = 0.27$ is adopted, but for some nearby
galaxies preferred distances are used.

\begin{figure*}
   \centering
\includegraphics[width=17.0cm]{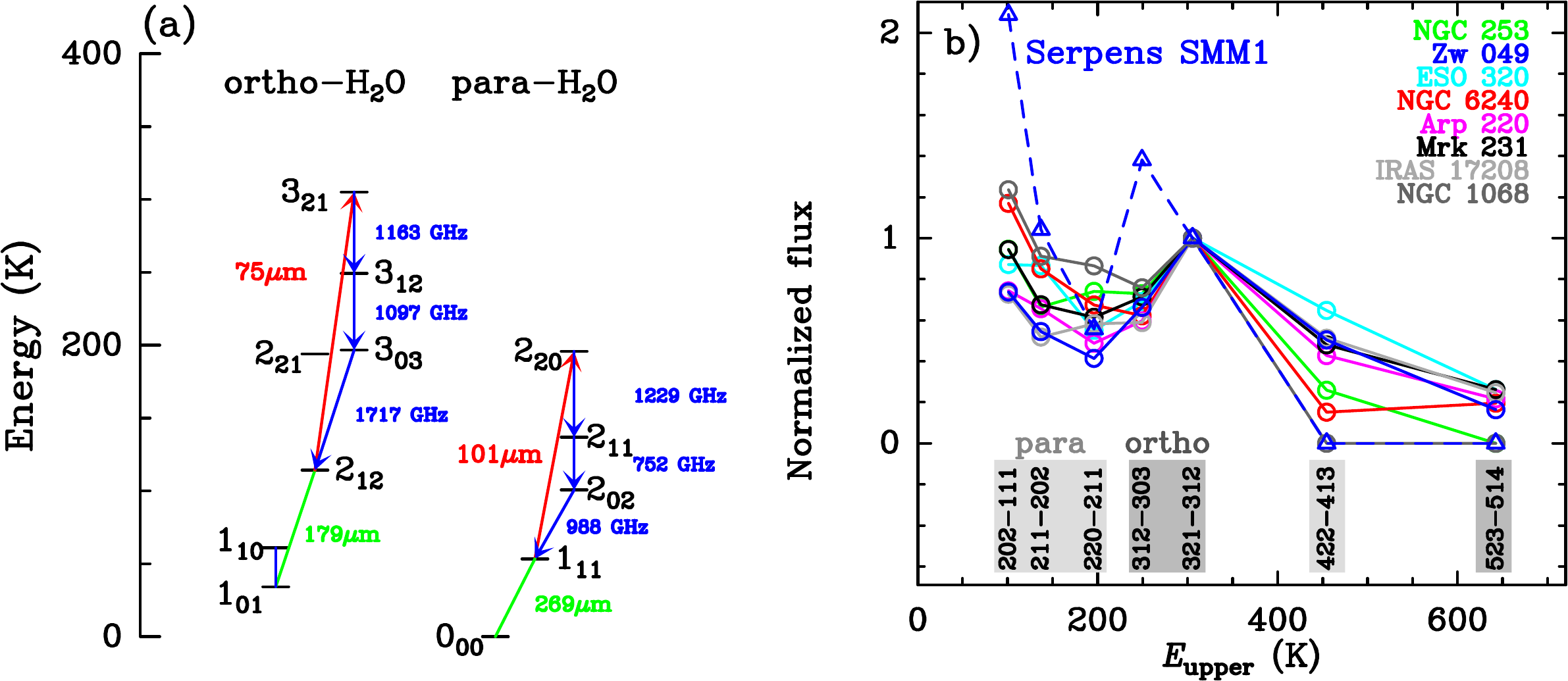}
\includegraphics[width=17.0cm]{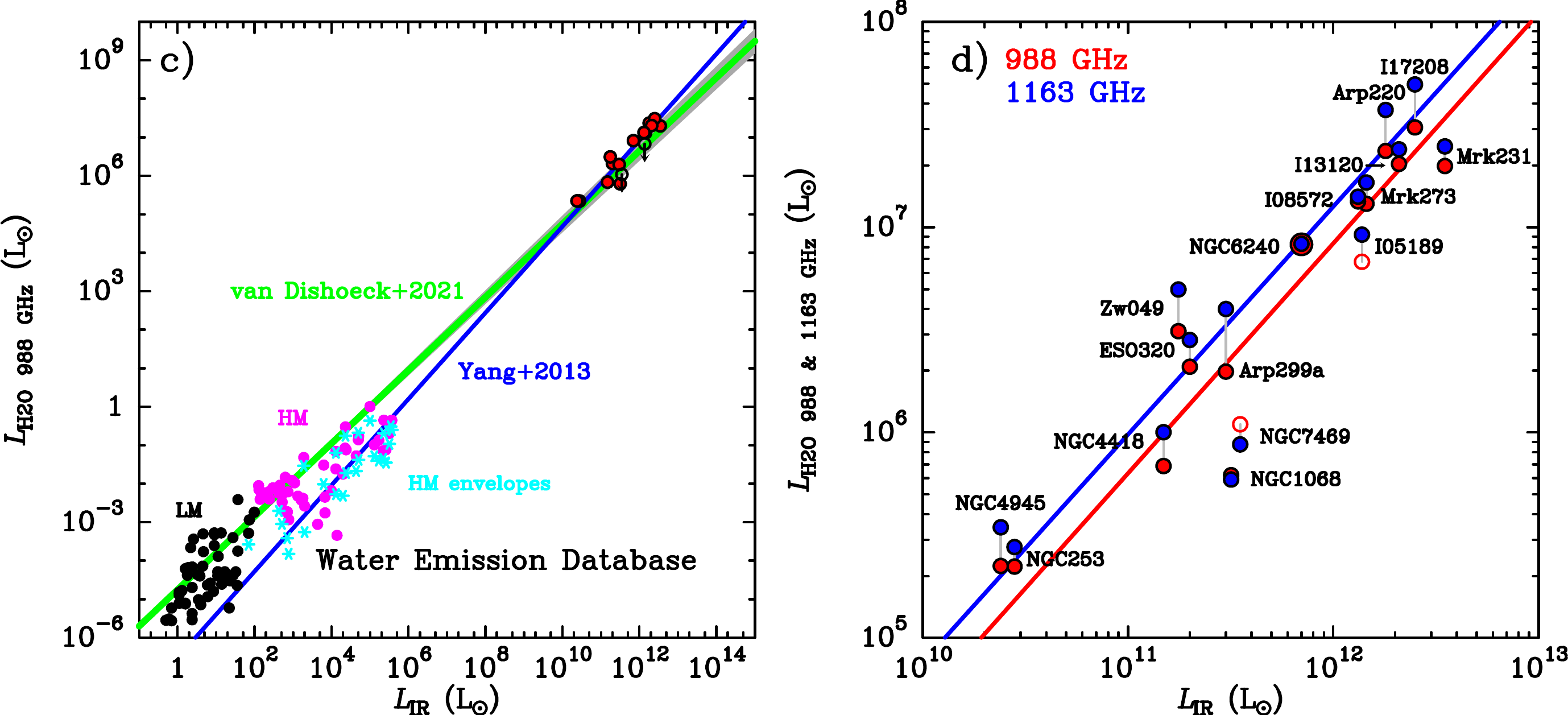}
\caption[]{
a) A simplified energy level diagram of ortho- and para-H$_2$O, illustrating
the radiative pumping mechanism of H$_2$O submillimeter emission
(Sect.~\ref{sec:diag}).
b) The Spectral Line Energy Distribution (SLED) of the H$_2$O
submillimeter lines, normalized to the flux of the $3_{21}\rightarrow3_{12}$
line, in 8 extragalactic sources where the pumping ortho-H$_2$O
  $3_{21}\leftarrow2_{12}$ line at 75\,$\mu$m has been observed, compared
  with the SLED in the LMSFR Serpens SMM1 \citep{goi12}. The flux of undetected
  lines is set to 0.
c) The luminosity
  of the H$_2$O\,$2_{02}\rightarrow1_{11}$\,988\,GHz line as a function of the
  source IR luminosity for both the extragalactic sources considered in this
  paper (red circles; open circles indicate $3\sigma$ upper limits)
  and the galactic low-mass and high-mass star-forming regions
  \citep[black and magenta circles, respectively, taken from the Water Emission
    Database;][]{dut22}\footnotemark.
  The light-blue symbols isolate
  the contribution to $L_{\mathrm{H_2O\,988\,GHz}}$ by the envelopes of HMSFRs.
  The green line shows the best-fit power-law function to all galactic and
  extragalactic sources found by \cite{vd21}, with an index of $0.95\pm0.02$,
  and the blue line shows the fit found to only (but all) the extragalactic
  sources by \cite{yan13}, with an index of $1.12$.
  d) Same as c) but zoomed-in on the extragalactic sources, with the luminosity
  of the H$_2$O\,$3_{21}\rightarrow3_{12}$\,1163\,GHz line added in blue. The
  red and blue lines show the fits by \cite{yan13}.
}
\label{sleds}%
\end{figure*}

\section{Diagnostics}
\label{sec:diag}

The simplified energy level diagram of Fig.~\ref{sleds}a illustrates the
basic mechanism of radiative pumping: if a 75\,$\mu$m photon emitted by dust
pumps an ortho-H$_2$O molecule from the $2_{12}$ level to the $3_{21}$ one,
it can relax via cascade down along the
\mbox{$3_{21}\rightarrow3_{12}\rightarrow3_{03}\rightarrow2_{12}$} ladder through
emission in the corresponding 1163, 1097, and 1717\,GHz lines. Likewise,
a pumping event in the para-H$_2$O $2_{20}\leftarrow1_{11}$\,101\,$\mu$m line
followed by cascade down along
the $2_{20}\rightarrow2_{11}\rightarrow2_{02}\rightarrow1_{11}$ ladder
generates H$_2$O emission in the 1229, 752, and 988\,GHz lines. These can
be denoted as ``radiative pumping cycles'', and always involve the loss of
a continuum photon in the 75 and 101\,$\mu$m pumping lines. Assuming that
the absorption is isotropic, the 75 and 101\,$\mu$m pumping lines
will be seen in absorption. On the other hand,
if H$_2$O is collisionally excited (e.g. in dense shocks),
the $3_{21}$ and $2_{20}$ levels will
be populated accordingly and subsequent spontaneous emission in the
$3_{21}\rightarrow2_{12}$ and $2_{20}\rightarrow1_{11}$ transitions will
generate emission, rather than absorption, in both the 75 and 101\,$\mu$m
lines. We can then establish the first qualitative criterion to distinguish
between radiative pumping and collisional excitation of the H$_2$O submm
lines assuming isotropy: in the former case, the 75 and 101\,$\mu$m lines are
expected in absorption, while in the latter case they are expected in emission.

Caveats on the above diagnostic are related to geometry effects.
Shock-excited gas could be in front of a strong continuum source 
that could ultimately generate line absorption at 75 and 101\,$\mu$m
in the direction of the observer.
However, since $T_{\mathrm{gas}}$ and $n_{\mathrm{H2}}$, which determine the
excitation in shocks, are fully decoupled from the properties of the
continuum source behind (and specifically $T_{\mathrm{gas}}>T_{\mathrm{dust}}$),
fine tuning of the shock model would be required to obtain the
absorption/emission pattern. On the other hand, lack of absorption in the
75 and 101\,$\mu$m lines does not fully preclude the radiative pumping mechanism
because the responsible far-IR field could be external without impinging on the
H$_2$O molecules from the back side (in the direction of the observer), and
hence would not produce line absorption \citep[Appendix A in][]{gon14}. 

A quantitative criterion for radiative pumping
arises from statistical equilibrium of the H$_2$O levels:
since the populations of the $3_{21}$ and $2_{20}$ levels
remain constant in time, and assuming that
their populations are exclusively determined by the above pumping cycles,
the net rate of absorption events in the $3_{21}\leftarrow2_{12}$\,75\,$\mu$m
line (hereafter $R^{\mathrm{abs}}_{\mathrm{75\,\mu m}}$) should be equal to the
net rate of emission events in the $3_{21}\rightarrow3_{12}$\,1163\,GHz line
($R^{\mathrm{ems}}_{\mathrm{1163\,GHz}}$). Likewise, the same equality 
holds for the para-H$_2$O pumping cycle,
$R^{\mathrm{abs}}_{\mathrm{101\,\mu m}}=R^{\mathrm{ems}}_{\mathrm{1229\,GHz}}$.
We calculate $R^{\mathrm{ems,abs}}_{\mathrm{line}}$ in the limit of optically
thin continuum emission at 75(101)\,$\mu$m and isotropic
line absorption/emission:
\begin{equation}
  R^{\mathrm{ems,abs}}_{\mathrm{line}} (\mathrm s^{-1}) =
  \pm \frac{L_{\mathrm{line}}}{E_{\mathrm{line}}} =
  \pm10^{-18}
  \frac{4\pi D_L^2\,F_{\mathrm{line}}}{(1+z)\, h \, c}
  \label{eq:rates}
\end{equation}
where $E_{\mathrm{line}}=h\nu_0$ is the energy of line photons,
$D_L$ is the luminosity distance, $z$ is the redshift,
$h$ is the Planck constant, $c$ is the speed of light, and
  $L_{\mathrm{line}}$ and $F_{\mathrm{line}}$ are the line luminosity in
  erg\,s$^{-1}$ and 
line flux in Jy\,km\,s$^{-1}$ above (emission, $+$ sign) or below
(absorption, $-$ sign) the continuum ($D_L$, $h$, and $c$ are in cgs
units).

Equation~(\ref{eq:rates}) holds if the H$_2$O\,75(101)\,$\mu$m line
is optically thick, provided that it remains {\it effectively}
optically thin and repeated absorption/re-emission events in the line
end with the photon either escaping from the source (with no contribution to
$R^{\mathrm{ems,abs}}_{\mathrm{line}}$) or generating a
submm cascade (contributing to both $R^{\mathrm{ems}}_{\mathrm{line}}$
and $R^{\mathrm{abs}}_{\mathrm{line}}$). However, if the far-IR continuum
optical depth becomes significant, thermal continuum photons emitted at
75(101)\,$\mu$m will have a higher chance, after multiple
line absorption/re-emission events, of being absorbed by dust grains
before escaping from the source,
generating absorption in the far-IR line with no submm line
counterpart. In very optically thick regions (with continuum
optical depths at 100\,$\mu$m $\tau_{100}>>1$ as found in a number of
(U)LIRGs) the emission in the submm lines will be partially extincted
and could even be observed in absorption.
As $\tau_{100}$ increases, additional radiative paths (de)populating the
$3_{21}$ and $2_{20}$ levels come into play, but the
overall general result is that the absorption fluxes of the
75 and 101\,$\mu$m surface tracers are increased relative to
the emission fluxes of the 1163 and 1229\,GHz volume tracers.
Therefore, in case of significant far-IR continuum optical depth effects,
the $R^{\mathrm{ems,abs}}_{\mathrm{line}}$ rates 
calculated in Eq.~(\ref{eq:rates}) are only apparent, and
we can more generally state that
\begin{eqnarray}
  R^{\mathrm{abs}}_{\mathrm{75\,\mu m}} & \ge & R^{\mathrm{ems}}_{\mathrm{1163\,GHz}}
  \label{eq:pump} \\
  R^{\mathrm{abs}}_{\mathrm{101\,\mu m}} & \ge & R^{\mathrm{ems}}_{\mathrm{1229\,GHz}}
  \nonumber 
\end{eqnarray}
are criteria for the radiative pumping mechanism in isotropic
conditions. Anisotropy effects of the exciting radiation field could favor
either $R^{\mathrm{abs}}_{\mathrm{75,101\,\mu m}}$ or
$R^{\mathrm{ems}}_{\mathrm{1163,1229\,GHz}}$, and would spread
their relative values for randomly oriented sources.

The ``base levels'' from which the radiative pumping cycles operate,
$2_{12}$ and $1_{11}$, must still be populated in some way.
They could be populated via absorption of dust-emitted photons in the
$2_{12}\leftarrow1_{01}$\,179\,$\mu$m and $1_{11}\leftarrow0_{00}$\,269\,$\mu$m
ground-state lines, but this mechanism may be inefficient in optically thin
$\tau_{100}<1$ sources where the dust emission at these wavelengths is weak.
At very high redshift, H$_2$O excitation from the ground-state can also
  be produced  by the cosmic microwave background \citep{rie22}.
Alternatively, the ``base levels'' could be excited through collisions in warm
and dense regions, so that collisional excitation of the low-excitation lines
combined with the radiative pumping mechanism would be
required to account for the submm emission \citep{gon14}.

\begin{figure*}
   \centering
\includegraphics[width=18.5cm]{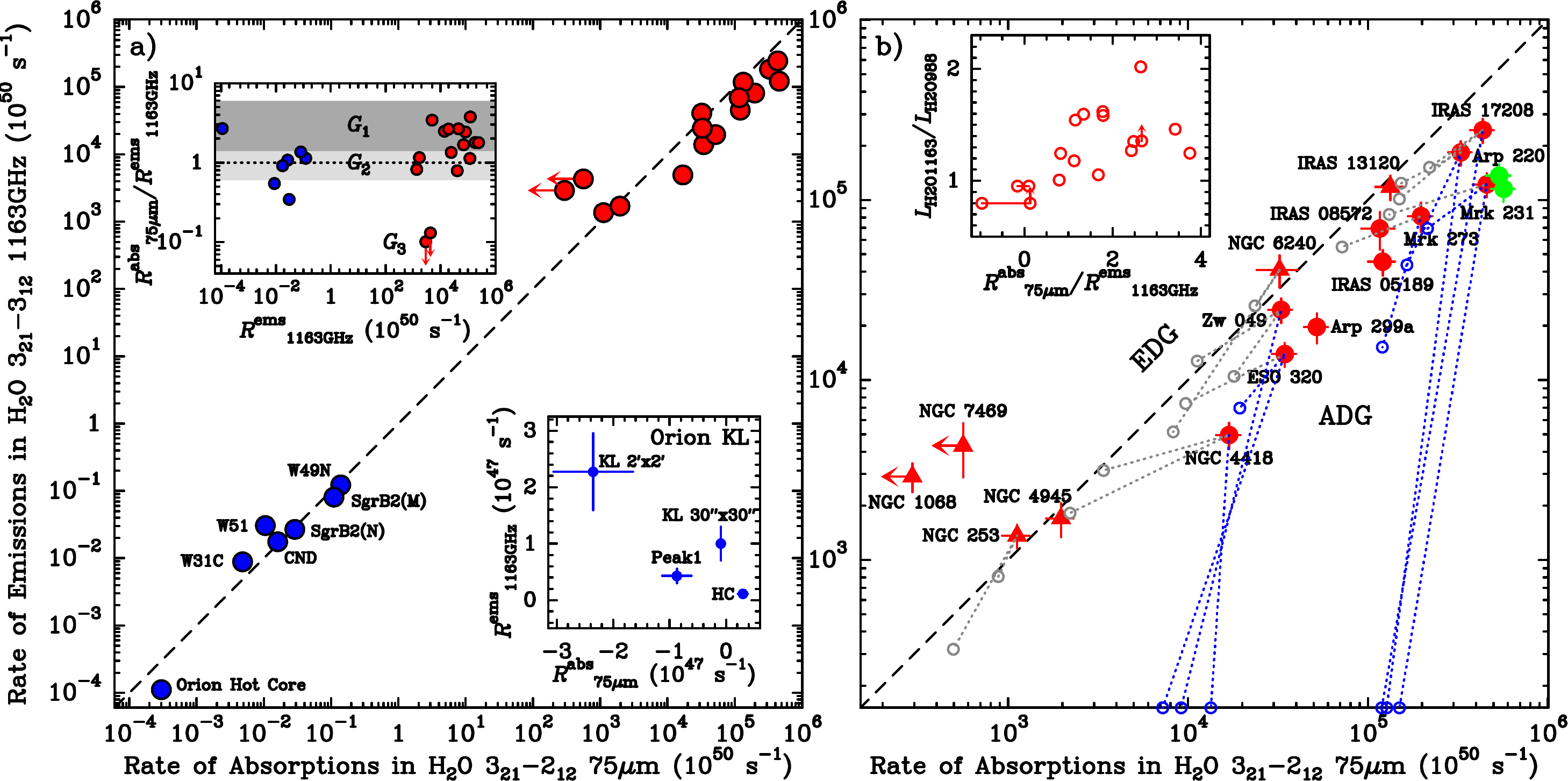}
\caption{
a) The quantitative diagnostic of the radiative pumping
mechanism, showing $R^{\mathrm{ems}}_{\mathrm{1163\,GHz}}$
as a function of
$R^{\mathrm{abs}}_{\mathrm{75\,\mu m}}$
for both the galactic and extragalactic samples. The dashed line indicates
$R^{\mathrm{abs}}_{\mathrm{75\,\mu m}}=R^{\mathrm{ems}}_{\mathrm{1163\,GHz}}$.
The lower insert shows these values in linear scale
for Orion KL, with negative numbers of $R^{\mathrm{abs}}_{\mathrm{75\,\mu m}}$
indicating emission in the 75\,$\mu$m line. The upper insert shows  
$r\equiv R^{\mathrm{abs}}_{\mathrm{75\,\mu m}}/R^{\mathrm{ems}}_{\mathrm{1163\,GHz}}$
versus $R^{\mathrm{ems}}_{\mathrm{1163\,GHz}}$,
with shading indicating the
  $G_1$, $G_2$, and $G_3$ groups defined according to the
  value of $r$ (Section~\ref{quantcrit}).
b) Same as a) but zoomed-in on the extragalactic sources. Circles and
  triangles indicate ADGs and EDGs, respectively (Section~\ref{samples}).
The green symbols show the same values but for the
$2_{02}\rightarrow1_{11}$ (emission) versus the $2_{20}\leftarrow1_{11}$
(absorption) in the two sources (Mrk\,231 and Arp\,220) where the 101\,$\mu$m
line has been observed. The open circles show the position in this
plane of the optically thin ($\tau_{100}<1$, in gray) and optically thick
($\tau_{100}>1$, in blue) model components used to fit the H$_2$O
emission/absorption (Appendix~\ref{app_sleds}). 
The insert shows the $3_{21}\rightarrow3_{12}$-to-$2_{02}\rightarrow1_{11}$
luminosity ratio versus
$R^{\mathrm{abs}}_{\mathrm{75\,\mu m}}/R^{\mathrm{ems}}_{\mathrm{1163\,GHz}}$,
and horizontal segments indicate the most likely ranges for the 2 sources
undetected in H$_2$O\,75\,$\mu$m.
}
\label{rates}%
\end{figure*}


\section{Results}
\label{results}

\subsection{The extragalactic and galactic samples}
\label{samples}

The crucial diagnostics for distinguishing between radiatively and
collisionally excited H$_2$O submm emission rely on sensitive observations at
75 and 101\,$\mu$m. We have applied them to all galaxies that
have PACS observations covering
\footnotetext{\url{https://katarzynadutkowska.github.io/WED/}}
the 75\,$\mu$m line, yielding a sample of 16 sources\footnote{With the
  exception of M82, which is not detected in the 75\,$\mu$m nor in the
  1163\,GHz line, and is therefore excluded from the sample.}.
Line fluxes and details of the flux derivations will be presented
in a forthcoming paper (J. Fischer et al., in prep.).
Unfortunately, the PACS
gap at $\sim100$\,$\mu$m precluded observations covering
the 101\,$\mu$m line
with {\it Herschel} in all galaxies but Mrk\,231 (due to its redshift).
This line, however, was detected with {\it ISO} in Arp\,220
\citep{fis14}.

The galactic sample includes 6 well-known HMSFRs where
the {\it Herschel}/PACS and SPIRE observations of the H$_2$O\,75\,$\mu$m
and 1163\,GHz lines have been carried out: W31C and W49N \citep{ger15},
W51 \citep{kar14},
Sgr B2(M) and (N) \citep{etx13}, and the Orion KL outflows \citep{goi15}.
In addition, the circumnuclear disk (CND) around Sgr A* \citep{goi13}
has been added.
The flux of the H$_2$O\,75\,$\mu$m line in this sample was extracted from the
central $3\times3$ spaxels of PACS ($\approx30''\times30''$). In Orion KL
we also measured the line flux in the direction of the Hot Core (hereafter HC,
a very bright far-IR continuum source with a size of $\sim10''$) within the
central spaxel ($\approx10''\times10''$), ``H$_2$ Peak 1'' shocked gas
position ($\approx30''\times30''$),
and Orion KL within a larger field of view (FoV of $\approx120''\times120''$).
The flux of the 1163\,GHz line was extracted from similar apertures.
The line profiles are displayed in Appendix~\ref{app_gal}.

Our extragalactic sample includes a variety of well-known local (U)LIRGs:
a QSO (Mrk\,231), AGNs such as IRAS~05189-2524, NGC\,1068, NGC\,7469,
NGC\,4945, IRAS~08572+3915, and NGC\,6240 
\citep[the latter with strong CO emission associated with shocks, see][]{mei13},
(U)LIRGs with a compact obscured nucleus
(NGC\,4418, Arp\,299a, Zw\,049.057, ESO\,320-G030, Mrk~273, Arp\,220,
IRAS\,17208-0014),
and starburst galaxies (NGC\,253 and IRAS\,13120-5453).
As delineated by the equivalent width of the OH\,65\,$\mu$m doublet
  \citep{fis14,gon15}, 6 galaxies (NGC\,1068, NGC\,7469, NGC\,253,
  NGC\,6240, IRAS\,13120-5453, and NGC\,4945) have
  overall weak far-IR molecular absorption features
  but usually strong emission in the atomic/ionic fine-structure lines
  ($\mathrm{EQW(OH\,65\,\mu m)}<20$\,km\,s$^{-1}$), 
  and will be referred to as ``emission-dominated galaxies'' (EDGs),
  while the rest (with $\mathrm{EQW(OH\,65\,\mu m)}>20$\,km\,s$^{-1}$)
  have stronger far-IR molecular absorption features and will be
referred to as ``absorption-dominated galaxies'' (ADGs).

The extragalactic submm SLEDs, with the H$_2$O line fluxes normalized to
that of the $3_{21}\rightarrow3_{12}$\,1163\,GHz line in Fig.~\ref{sleds}b,
show a rather common $U$-shaped pattern for the low-excitation
($E_{\mathrm{up}}\lesssim300$\,K) lines:
the 1163\,GHz line is usually the strongest line followed by the
988\,GHz line, except in NGC\,1068, NGC\,6240, and probably NGC\,7469
where this sequence is interchanged.
This strongly suggests that there is a common dominant excitation mechanism
in all sources, but with variations. By contrast, the submm SLED in the
LMSFR Serpens SMM1
\citep[a template of collisional excitation in dense shocked gas, e.g.][]{goi12}
looks different, with the fluxes of the para- and
ortho- lines sharply decreasing with increasing $E_{\mathrm{up}}$.

As shown in Fig.~\ref{sleds}c, most of the targets lie close to the
$L_{\mathrm{H_2O-988\,GHz}}-L_{\mathrm{IR}}$ correlations found by
\cite{yan13} (for extragalactic sources) and \cite{vd21} (for both galactic and
extragalactic sources), and can thus be considered a representative subsample
of (mostly) (U)LIRGs.
A closer inspection (Fig.~\ref{sleds}d) reveals that NGC\,7469 and NGC\,1068
have important deficits in the emission of both the
H$_2$O\,988\,GHz ($2_{02}-1_{11}$) and mostly the 1163\,GHz ($3_{21}-3_{12}$)
lines, relative to the best global $L_{\mathrm{H_2O}}-L_{\mathrm{IR}}$ fit
by \cite{yan13}, with departures of $\gtrsim3$. Together with NGC\,6240,
these sources also have the lowest 1163\,GHz/988\,GHz flux ratio.

\subsection{The qualitative criterion}
\label{qualitative}

Of the 16 galaxies, 14 (87\%) show the H$_2$O\,75\,$\mu$m
far-IR line in absorption (and the 101\,$\mu$m line in Mrk\,231 and
Arp\,220). 
The other two sources, NGC\,1068 and NGC\,7469, are ambiguous because
the 75\,$\mu$m is not detected at the $3\sigma$ level.
The 75\,$\mu$m spectrum of NGC\,1068 is displayed in
Appendix~\ref{app_outl}, showing hints
of a P Cygni profile that nearly cancels out the 
blueshifted negative flux and the redshifted positive flux.

In galactic HMSFRs,
the H$_2$O\,75\,$\mu$m line is observed in absorption in all sources
except Orion KL (Fig.~\ref{highmass}). 
In Orion KL, the nature of the line depends on the specific observed
region and FoV: the 75\,$\mu$m line is observed in absorption
towards the HC within the central PACS spaxel, but
shows a P Cygni profile or is observed in emission for larger FoVs
and towards Peak 1
(Fig.~\ref{highmass}).

By contrast, observations of low-mass Class 0 and I and intermediate mass
protostars where H$_2$O is
collisionally excited in shocks show the H$_2$O far-IR lines (including the
75\,$\mu$m line when observed, and with the exception of the 179\,$\mu$m
line in some sources)
in emission \citep{her12,goi12,kar14b,kar18,mat15}.

\subsection{The quantitative criterion}
\label{quantcrit}

Figure~\ref{rates}a shows $R^{\mathrm{ems}}_{\mathrm{1163\,GHz}}$
as a function of $R^{\mathrm{abs}}_{\mathrm{75\,\mu m}}$ for both samples,
and Figure~\ref{rates}b zooms in on the extragalactic sources.
We categorize the galaxies in this plane into 3 groups
according to the
  $r\equiv R^{\mathrm{abs}}_{\mathrm{75\,\mu m}}/R^{\mathrm{ems}}_{\mathrm{1163\,GHz}}$
  ratio (see also the upper insert of Fig.~\ref{rates}a):
  $G_1$: 9 galaxies show $r>1.4$;
$G_2$: 5 sources lie close, within the calibration uncertainties, to the
$R^{\mathrm{abs}}_{\mathrm{75\,\mu m}}=R^{\mathrm{ems}}_{\mathrm{1163\,GHz}}$ line
($0.6<r<1.4$);
$G_3$: 2 galaxies, NGC\,1068 and NGC\,7469, have only upper limits in
$R^{\mathrm{abs}}_{\mathrm{75\,\mu m}}$ and show much higher
$R^{\mathrm{ems}}_{\mathrm{1163\,GHz}}$. 
With the exception of Zw\,049.057, all ADGs belong to $G_1$, so that
the plane of Fig.~\ref{rates} can be used to discriminate between ADGs and EDGs
from the observation of only two lines.
On the other hand, most of the galactic HMSFRs of our sample
except Orion KL lie close to the
$R^{\mathrm{abs}}_{\mathrm{75\,\mu m}}/R^{\mathrm{ems}}_{\mathrm{1163\,GHz}}=1$ line,
with values ranging from 0.34 (W51) to 1.4 (SgrB2(M)),
and can thus be classified as belonging to $G_2$. Orion HC
shows an excess of $R^{\mathrm{abs}}_{\mathrm{75\,\mu m}}$ similar to
$G_1$ sources, but once the
FoV increases (KL $30''\times30''$ and $2'\times2'$) or towards Peak 1,
$R^{\mathrm{abs}}_{\mathrm{75\,\mu m}}$ becomes negative (i.e. it is observed
in emission, lower insert in Fig~\ref{rates}a).

Overall, most of the (U)LIRGs and HMSFRs observed in the H$_2$O\,75\,$\mu$m
line (nearly) follow the prediction of Eq.~(\ref{eq:pump}) over 9 orders
of magnitude in $R^{\mathrm{ems}}_{\mathrm{1163\,GHz}}$, indicating
that far-IR pumping is the essential ingredient to understand
the H$_2$O submm emission and specifically it
dominates the observed H$_2$O\,$3_{21}-3_{12}$\,1163\,GHz emission
(and thus also the $3_{12}-3_{03}$\,1097\,GHz emission) in the bulk of
sources. In some galaxies and HMSFRs, there is an excess of
submm emission over far-IR absorption 
($R^{\mathrm{abs}}_{\mathrm{75\,\mu m}}/R^{\mathrm{ems}}_{\mathrm{1163\,GHz}}=0.34-1$),
which could reflect either a contribution to the submm line flux by
shock-excited H$_2$O or specific geometrical effects.

Multicomponent model fits to the emission/absorption fluxes of both the
SPIRE and the PACS H$_2$O lines, and including the far-IR
spectral energy distribution (SED),
have been performed for most galaxies of our sample following the
approach described in \cite{gon21} 
(these will be presented in a forthcoming paper). Classifying the best-fit
model components of each galaxy as ``optically thin'' if the optical depth
of the continuum at 100\,$\mu$m is lower than 1 ($\tau_{100}<1$), and
``optically thick'' if $\tau_{100}\ge1$, Fig.~\ref{rates}b
shows that the $\tau_{100}<1$ components (in gray) lie close to the
$R^{\mathrm{abs}}_{\mathrm{75\,\mu m}}=R^{\mathrm{ems}}_{\mathrm{1163\,GHz}}$ line
as expected, while the optically thick components (in blue) predict 
weak emission in the 1163\,GHz line, and even in some extreme nuclei in
absorption.  The ADGs have
$R^{\mathrm{abs}}_{\mathrm{75\,\mu m}}>R^{\mathrm{ems}}_{\mathrm{1163\,GHz}}$
because of the presence of buried components that generate strong absorption in
the H$_2$O\,75\,$\mu$m but little emission in the submm lines due to
high continuum brightness and extinction in the submm. The submm lines are
in contrast formed in optically thin/moderately thick and more extended regions,
likely the $0.1-0.5$\,kpc circumnuclear disks of (U)LIRGs.

In Appendix~\ref{app_sleds} we compare the observed submm SLEDs of 8
sample galaxies and the best-fit predictions by the multicomponent models,
generally showing good agreement with a minimum emission line flux in
the $2_{20}-2_{11}$\,1229\,GHz line. We also show that in some EDGs
(NGC\,6240 and NGC\,253) collisional excitation of the 
base levels is required to make the radiative pumping operational,
and finally discuss in Appendix~\ref{app_outl} whether the cases of NGC\,7469
and NGC\,1068 represent sources where H$_2$O is shock-excited, or
whether geometrical effects can account for the lack of H$_2$O\,75\,$\mu$m
absorption.

\section{Discussion and conclusions}
\label{disc}

The preponderance of H$_2$O\,75\,$\mu$m absorption is difficult to reconcile
with collisional excitation in shock cavities devoid of significant amounts
of warm dust. An alternative is to assume that
the H$_2$O\,1163\,GHz line is generated in an ensemble of shock cavities,
where the H$_2$O\,75\,$\mu$m line is also generated in emission,
while a spatially separated and unrelated $\tau_{100}>>1$ component produces
the 75\,$\mu$m absorption (almost always stronger than the emission to
generate net absorption in the line). 
However, this scenario does not explain the correlation between
$R^{\mathrm{abs}}_{\mathrm{75\,\mu m}}$ and $R^{\mathrm{ems}}_{\mathrm{1163\,GHz}}$
as the two lines are generated in different regions and by different mechanisms.
Even if $\dot{E}$ (mechanical) and $L_{\mathrm{IR}}$ are related, the specific
lines respond in different ways to the far-IR and mechanical feedback.
The case of Orion KL illustrates this point: large FoVs show emission
in the 75\,$\mu$m line far above the absorption towards the HC
(lower insert in Fig.~\ref{rates}a).

The following interpretation of the $L_{\mathrm{H2O\,988\,GHz}}-L_{\mathrm{IR}}$
correlation stems from the present analysis.
In Fig.~\ref{sleds}c, we have increased the dynamic range of
the correlation found by \cite{yan13} for extragalactic sources
to the luminosities characteristic of galactic sources, assuming
that the ``radiative pumping slope'' of 1.12 holds. Then, a number of
HMSFRs and basically all LMSFRs, where H$_2$O is shock-excited, lie well
above this line. Using the spectral decomposition of the
H$_2$O\,988\,GHz spectra in HMSFRs carried out by \cite{tak13} and
\cite{sanjose16}, we have plotted in Fig.~\ref{sleds}c the H$_2$O\,988\,GHz
luminosities due to only the massive envelopes of HMSFRs (light-blue symbols),
finding a good match with the extended extragalactic correlation. 
We thus propose that these galactic massive envelopes play a role similar
to the circumnuclear disks in galaxies (including the CND of the Milky Way),
where the H$_2$O submm emission is radiatively pumped\footnote{In this
framework, the slope of 1.1 could be due to the increasing linewidth
$\Delta v$ with $L_{\mathrm{IR}}$, as for the optically thick 75 and
101\,$\mu$m lines $R^{\mathrm{abs}}_{\mathrm{75-101\,\mu m}}\propto \Delta v$.
If $\Delta v$ increases by a factor $\sim5$ as $L_{\mathrm{IR}}$ increases
by 7 dex, the index of the correlation will be $\sim1+(\log_{10}5)/7=1.1$.}
-combined at least in some cases with
collisional excitation of the low-energy levels.
Galactic sources located significantly above the blue line in Fig.~\ref{sleds}c
would indicate dominance of excitation by shocks.
The location of the galactic HMSFRs in Fig.~\ref{rates}a resemble the
starburst galaxies NGC\,253 and IRAS\,13120-5453, lacking an 
obscured nucleus that in its prominent form
is a unique feature of extragalactic sources.

While we expect that the para-H$_2$O 101\,$\mu$m pumping cycle is as
important as the ortho-H$_2$O 75\,$\mu$m one, the low-excitation
$1_{11}$ and $2_{02}$
levels will be more affected by collisions in warm/dense regions than the
$3_{21}$ one; therefore the 988\,GHz line has less ability than the 1163\,GHz
line to discriminate between the two mechanisms.

\begin{acknowledgements}
  We thank the referee for helpful comments that improved the clarity
    of the manuscript.
 EG-A is a Research Associate at the Harvard-Smithsonian
Center for Astrophysics, and thanks the Spanish MICINN
for support under project PID2019-105552RB-C41.
JRG thanks the Spanish MCINN for funding support under grant
PID2019-106110GB-I00.
JF and KPS gratefully acknowledge support
through NASA grant NNH17ZD001N-ADAP.
C.Y. acknowledges support from ERC Advanced Grant 789410.
MPS acknowledges support from the Comunidad de Madrid through the Atracci\'on
de Talento Investigador Grant 2018-T1/TIC-11035 and PID2019-105423GA-I00
(MCIU/AEI/FEDER,UE).
PACS was developed by a consortium of institutes
led by MPE (Germany) and including UVIE (Austria); KU Leuven, CSL, IMEC
(Belgium); CEA, LAM (France); MPIA (Germany); 
INAFIFSI/OAA/OAP/OAT, LENS, SISSA (Italy); IAC (Spain). This development
has been supported by the funding agencies BMVIT (Austria), ESA-PRODEX
(Belgium), CEA/CNES (France), DLR (Germany), ASI/INAF (Italy), and
CICYT/MCYT (Spain).
SPIRE was developed by a consortium of institutes led by Cardiff University
(UK) and including Univ. Lethbridge (Canada); NAOC (China); CEA, LAM (France);
IFSI, Univ. Padua (Italy); IAC (Spain); Stockholm Observatory (Sweden);
Imperial College London, RAL, UCL-MSSL, UKATC, Univ. Sussex (UK);
and Caltech, JPL, NHSC, Univ.Colorado (USA). This development has been
supported by national funding agencies:
CSA (Canada); NAOC (China); CEA, CNES, CNRS (France); ASI (Italy);
MCINN (Spain); SNSB (Sweden); STFC, UKSA (UK); and NASA (USA).
\end{acknowledgements}

\bibliographystyle{aa}
\bibliography{refs}

\begin{appendix}

\section{The H$_2$O\,75\,$\mu$m profiles in galactic high-mass star-forming
regions}
\label{app_gal}

\begin{figure}[h]
   \centering
\includegraphics[width=9.0cm]{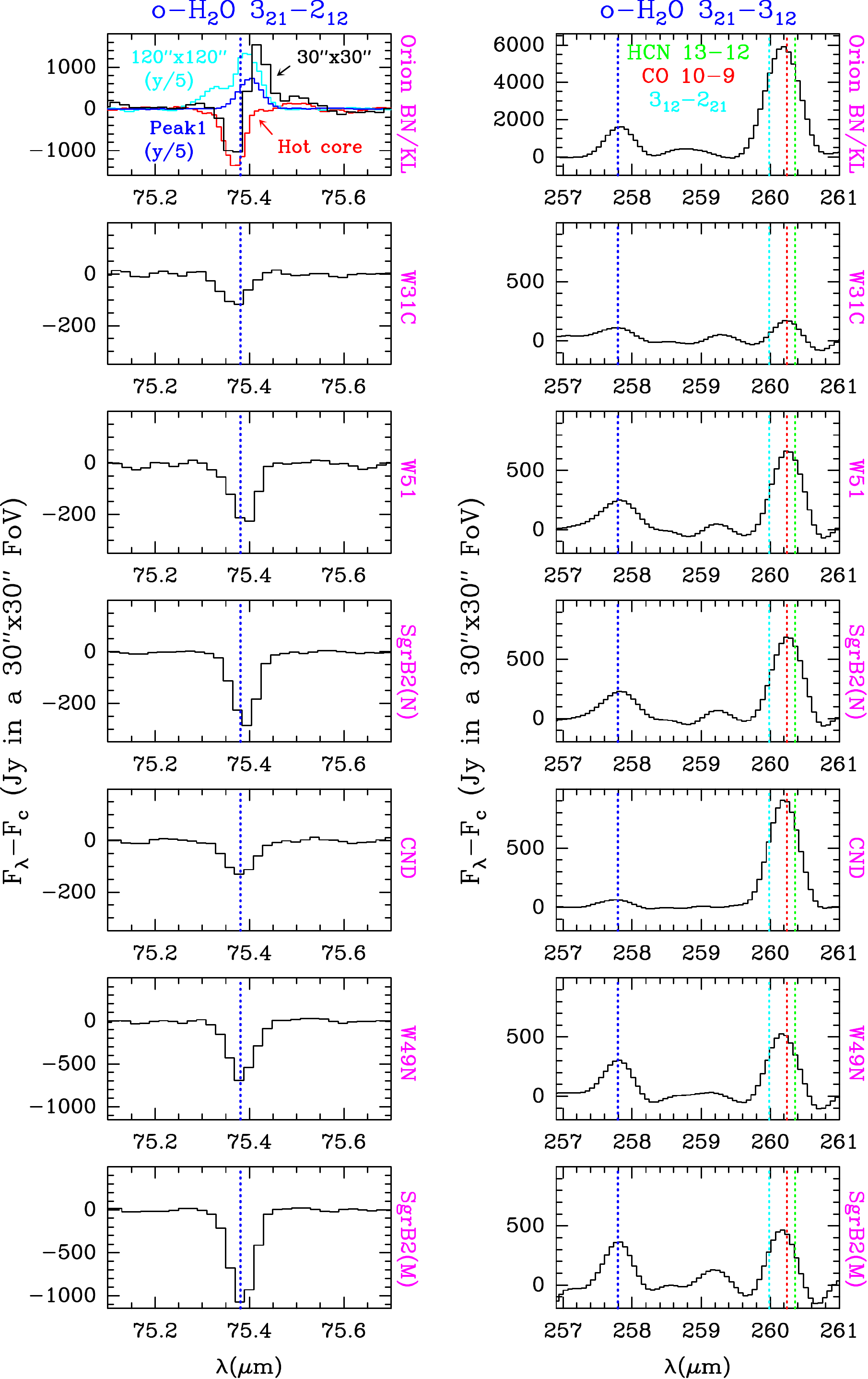}
\caption{The H$_2$O $3_{21}-2_{12}$\,75\,$\mu$m (left) and
  H$_2$O $3_{21}-3_{12}$\,1163\,GHz (right) profiles in 7 galactic high-mass
  star-forming regions, as observed with {\it Herschel}/PACS and SPIRE.
  The apodized SPIRE profiles also include the blend of CO\,10-9 (which
  dominates the observed emission),
  H$_2$O $3_{12}-2_{21}$, and HCN\,13-12. The PACS spectra correspond to
  the central $3\times3$ spaxels ($\approx30''\times30''$). In Orion KL,
  the H$_2$O\,75\,$\mu$m line is observed in absorption towards the Hot Core
  as seen within the central PACS spaxel ($\approx10''\times10''$), but
  shows P Cygni and emission profiles for larger field-of-views and towards
  Peak 1 (upper-left panel; see also the insert in Fig.~\ref{rates}a).
}
\label{highmass}%
\end{figure}

Figure~\ref{highmass} shows the H$_2$O\,75\,$\mu$m and 1163\,GHz profiles
in our sample of galactic high-mass star-forming regions.

\FloatBarrier

\section{The submillimeter SLEDs}
\label{app_sleds}

\begin{figure*}
   \centering
\includegraphics[width=15.0cm]{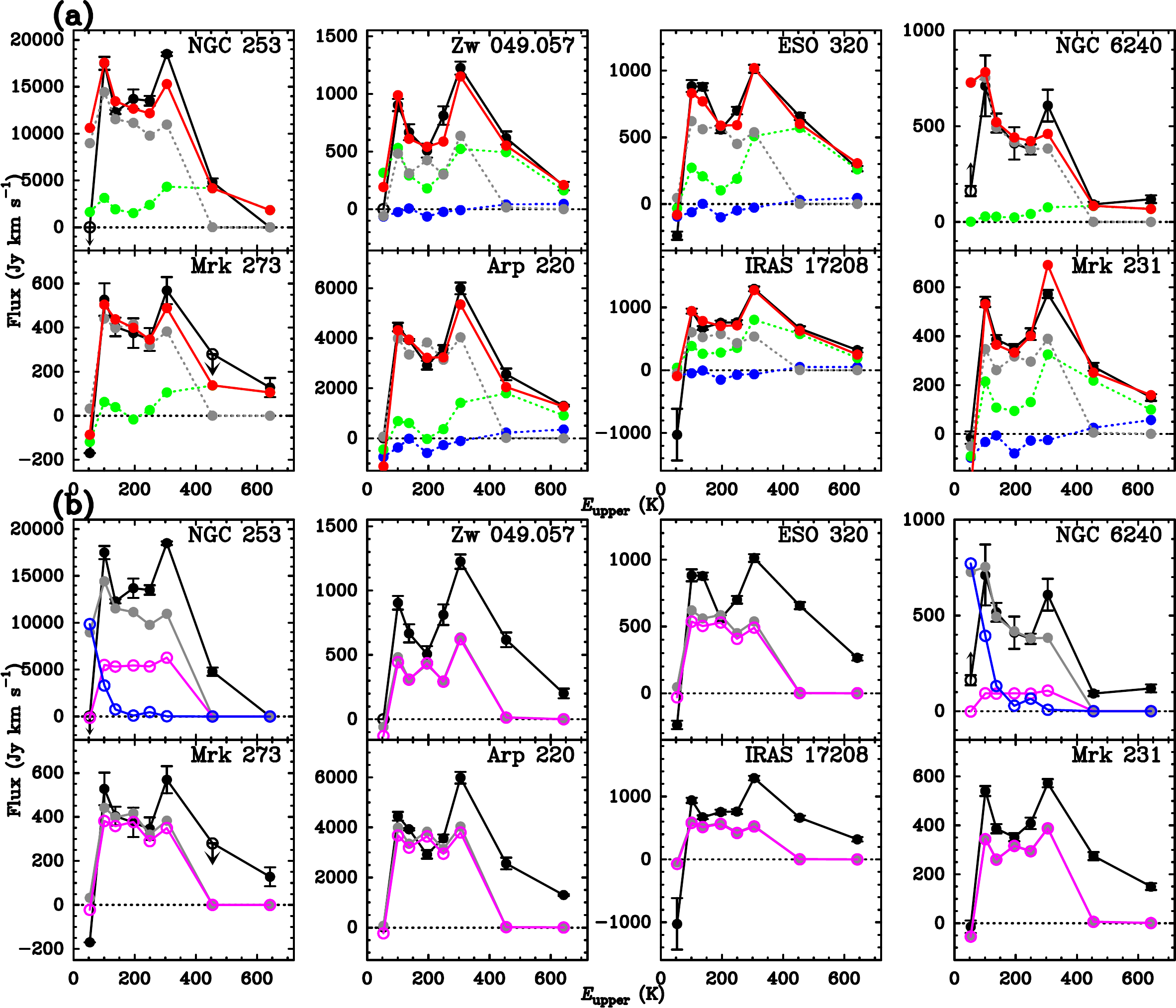}
\caption{a) Fits to the H$_2$O submillimeter emission for 8 local
  extragalactic sources analyzed in this paper, including all modeling
  components. Black circles with errorbars show the observed fluxes.
  Blue, green, and gray circles indicate the contributions to the model fit
  by the core, disk, and envelope components, respectively, and red is total. 
  b) The model component that dominates the emission of the H$_2$O
  low-excitation submillimeter lines (the envelope component) is shown
  with gray circles and lines.
  For comparison, magenta symbols and lines show the modeling results for
  these components obtained when the collisional excitation is quenched,
  and blue symbols (for NGC 253 and NGC 6240) indicate the results obtained
  when the far-IR pumping is quenched (but collisional excitation is included).
}
\label{fits}%
\end{figure*}

Figure~\ref{fits}a compares the observed submm SLEDs with the best-fit
predictions (in red) by the multicomponent models applied to 8 sample galaxies,
4 (sub)LIRGs and 4 ULIRGs \citep[see also][for similar models applied
  to Zw~049-057 and Arp~299a]{fal15,fal17}.
In these models, collisional excitation was only roughly simulated with
a single value of $T_{\mathrm{gas}}=150$\,K and 3 values of the density
($n(\mathrm{H_2}))$: $1.7\times10^4$, $5\times10^4$, and
$1.5\times10^5$\,cm$^{-3}$ \citep{gon21}. Only in the case of NGC\,6240,
the highest density was favored for the component that dominates the
H$_2$O submm emission. Although with some discrepancies, the overall SLEDs
are naturally reproduced with far-IR radiation fields that also account
for the PACS lines and the SED.

For the model components in Fig.~\ref{fits}a, we use the same nomenclature
as in \cite{gon21}:
blue, green, and gray circles indicate the contributions to the model fit
by the ``core'', ``disk'', and ``envelope'' components, respectively,
and red is total.
Schematically, the high-excitation absorption lines are formed in the
far-IR photosphere of the optically thick and very warm cores;
the medium-excitation absorption and emission lines have contributions
from the disk components, and the envelopes dominate the H$_2$O submm line
emission and also generate some absorption in the low-excitation far-IR lines
(such as the 75\,$\mu$m line).

A general characteristic of the H$_2$O submm SLEDs of galaxies is their
$U-$shape, shown also in Fig.~\ref{sleds}b.
It is instructive to understand that, according to our non-local models,
the minimum emission line flux in most cases, which corresponds
to the para-H$_2$O\,$2_{20}-2_{11}$\,1229\,GHz line, is due to competing
{\it absorption} of the continuum by the line in the optically thicker core and
disk components, rather than an intrinsic weakness of the 1229\,GHz line
emission from the optically thin envelope.  
The main reason that the absorption in the 1229\,GHz
line is the strongest is that it has an $A_{ul}-$Einstein
coefficient $2.7\times$
higher than that of the $2_{11}-2_{02}$\,752\,GHz line.
Close to the source surface, this has the effect of reducing
the excitation temperature ($T_{\mathrm{ex}}$) of the 1229\,GHz line, which will
be thus more prone to absorb the brighter (due to its also higher frequency)
continuum behind.
The same effect does not occur with the
ortho-H$_2$O\,$3_{21}-3_{12}$\,1163\,GHz line, because its $A_{ul}$ is only
$1.2\times$ higher than $A_{ul}(3_{12}-3_{03})+A_{ul}(3_{12}-2_{21})$, and thus
the 1163\,GHz line remains relatively excited (and even with 
supra-thermal excitation in some models) close to the surface.
The described absorption effect specific to the 1229\,GHz line in ADGs
is indeed clearly seen in our data: in Fig.~\ref{sleds}b, Arp\,220
(with a prominent core) shows one of the lowest (relative) 1229\,GHz
fluxes, while the EDGs NGC\,1068 and NGC\,253 do not show that dip. 
In summary, while the flux of the H$_2$O submm emission is dominated by the
envelopes, the cores and disks modulate the line ratios generating the
SLED $U-$shaped pattern. 

  In contrast with the nearly flat submm SLEDs for the envelopes predicted
  by the radiative pumping scenario (in absence of important collisional
  excitation of the low-excitation lines, see below), the SLED of Serpens
  SMM1\footnote{
    The Serpens SMM1 line fluxes were taken from Table A.1 of \citet{goi12},
    with the exception of a few fluxes that appeared with typographical errors.
    The H$_2$O $2_{20}-2_{11}$ and $2_{21}-1_{10}$ lines should have been listed
    as $1.46\times10^{-16}$ and $1.35\times10^{-15}$\,W\,m$^{-2}$ instead of
    $2.92\times10^{-17}$ and $1.35\times10^{-16}$\,W\,m$^{-2}$, respectively.}
  shows a strong decline of the fluxes of the para- and ortho- lines
  with increasing $E_{\mathrm{up}}$ (Fig.~\ref{sleds}b). Here, the dip in
  the para-H$_2$O\,$2_{20}-2_{11}$\,1229\,GHz line has nothing to do with
  the absorption of any continuum, but with the difficulty of exciting the
  $2_{20}$ level without a continuum source emitting at 101\,$\mu$m. A
  characteristic of H$_2$O shock excitation in dense gas is that the
  ortho $3_{12}-3_{03}$\,1097\,GHz line is expected to be stronger than
  the $3_{21}-3_{12}$\,1163\,GHz line, as opposed to the situation when
  radiative pumping dominates.

For each of the 8 sources, Fig.~\ref{fits}b shows the predictions for only
the envelope component that dominates the H$_2$O submm emission (gray symbols).
To better understand the relative roles of collisional and radiative pumping,
we have generated exactly the same models but quenching the collisional
rates (in magenta). While in the 4 ULIRGs results remain the same, meaning
that collisional excitation has little effect on the H$_2$O excitation,
the quenching of collisions has a strong effect in some sub-ULIRGs, specifically
in NGC~253 and NGC~6240. In these EDGs, we also generated the same models
but quenching the pumping radiation field and keeping the collisional excitation
(in blue), with the result that the line fluxes dropped dramatically
for all except the $1_{11}-0_{00}$ and $2_{02}-1_{11}$ lines.
This indicates that collisional excitation is required to populate the
``base levels'' ($2_{12}$ and $1_{11}$)
from which the pumping cycles operate \citep{gon14}. In these
sources, it is the combination of collisional excitation of the base levels and
radiative pumping from them that generates the H$_2$O submm emission.
These results also suggest that in the highest luminosity galaxies,
collisional excitation of the base levels tends to be less important.

\section{Outliers (NGC\,1068 and NGC\,7469): collisional excitation or
  geometrical effects?}
\label{app_outl}

NGC\,1068 and NGC\,7469 are ambiguous sources, because the H$_2$O\,75\,$\mu$m
line is neither detected in absorption, as expected in the scenario of
radiative pumping, nor in emission, as expected in case of shock
excitation. As shown in Fig.~\ref{sleds}d, both sources are underluminous in
both H$_2$O\,1163\,GHz and 988\,GHz lines relative to $L_{\mathrm{IR}}$,
with a stronger deficit in the 1163\,GHz line. These galaxies are thus not
representative of the bulk of (U)LIRGs observed with SPIRE and fitted by
\cite{yan13}, but may well represent a population of low-luminosity AGNs
($\lesssim10^{11}\,L_{\odot}$) with starburst rings under-sampled by
{\it Herschel}/PACS and SPIRE.

\begin{figure}
   \centering
\includegraphics[width=8.cm]{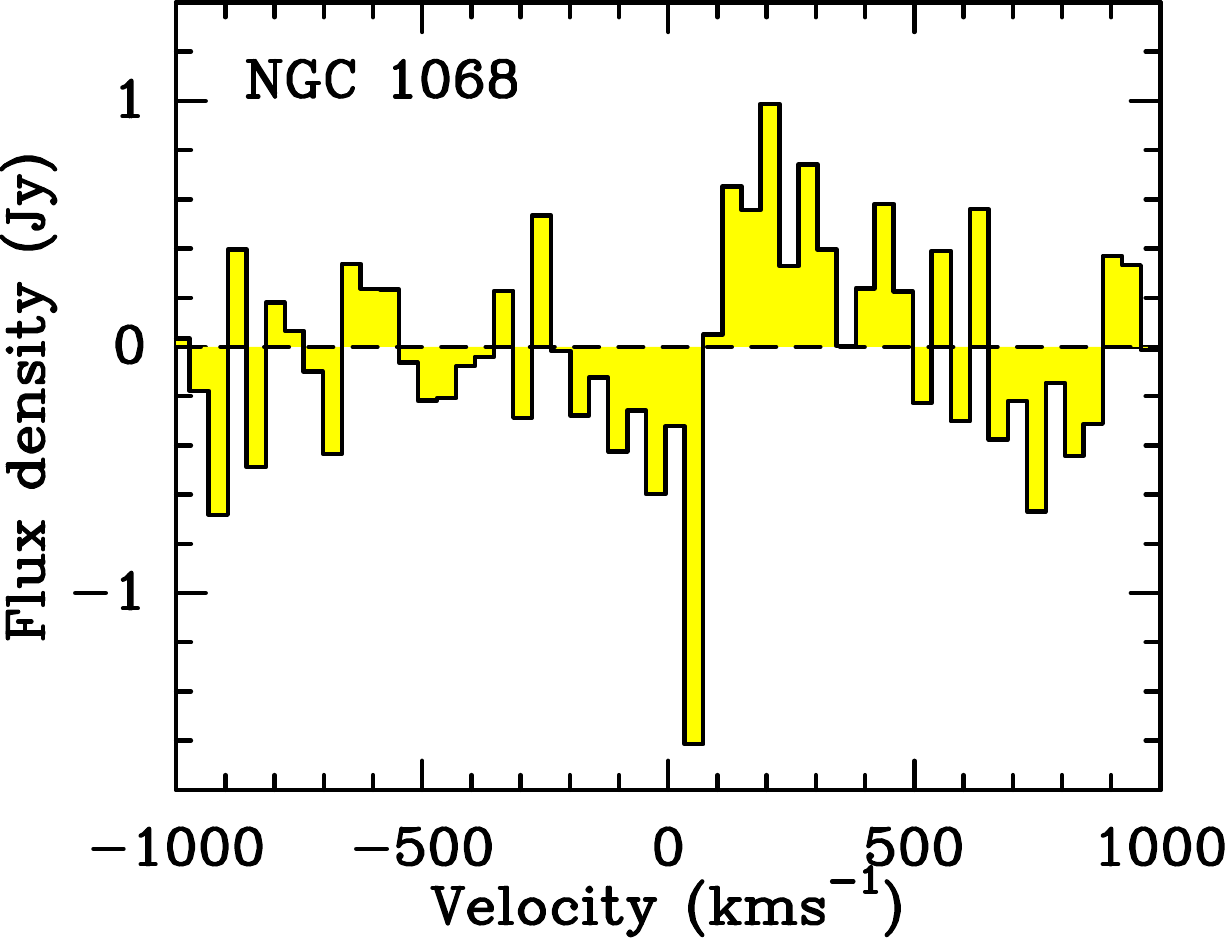}
  \caption{The H$_2$O\,$3_{21}-2_{12}$\,75\,$\mu$m continuum-subtracted
    spectrum in NGC\,1068. 
}
\label{ngc1068}%
\end{figure}

The PACS continuum-subtracted spectra around the
H$_2$O\,$3_{21}-2_{12}$\,75\,$\mu$m line in NGC\,1068 and NGC\,7469 are 
shown in Fig.~\ref{ngc1068}. The net flux of the line in both sources
is below $3\sigma$ ($42\pm64$ and $73\pm30$\,Jy\,km\,s$^{-1}$), but
hints of a P Cygni profile are seen in the spectrum of NGC\,1068
with the absorption and emission features nearly cancelling each
other. P Cygni profiles with similar absorption and emission fluxes
are characteristic of radiatively excited lines. We note that the redshifted
emission component will also contribute to the 1163\,GHz flux via radiative
pumping without generating 75\,$\mu$m absorption because the emitting gas
is behind the continuum source. This is one example of anisotropy alluded to in
Section~\ref{sec:diag}. However, the measured flux in the 75\,$\mu$m
blueshifted absorption component falls too short to account for the
observed strong submm emission in the 1163\,GHz line.

\cite{spi12} reported and analyzed the H$_2$O submm emission in NGC\,1068,
including the detected PACS lines (all in emission). The H$_2$O 
emission was found to be dominated by the CND around the AGN.
Collisional excitation of H$_2$O is important in the CND,
as indicated by e.g. the relatively strong emission of the
H$_2$O\,$1_{11}-0_{00}$\,1113\,GHz
line, and by a 1163-to-988 flux ratio lower than 1 (Fig.~\ref{sleds}b).
However, the proposed LVG models that ignored radiative
pumping yielded $T_{\mathrm{gas}}\sim40$\,K for the H$_2$O in the CND that
did not match the conditions inferred from the CO lines
\citep[$T_{\mathrm{gas}}\sim170-570$\,K,][]{hai12}. 
An alternative model was then explored by \cite{gon14}, which included
the radiative pumping effect by the dust mixed with H$_2$O and also by an
external far-IR radiation field that was anisotropic, that is, it did
not impinge onto the clumps in the direction
of the observer and thus did not produce absorption of the far-IR lines
in the direction of the Earth.
On the contrary, the external field produced emission in the far-IR lines,
which would nearly cancel the absorption by the internal field in the
case of the 75\,$\mu$m line. A similar situation could take place in
the Sy 1.2 galaxy NGC\,7469.

\end{appendix}

\end{document}